\documentclass[aps,amsmath,nofootinbib,superscriptaddress]{revtex4}

\usepackage{epsfig}
\usepackage{graphicx}

\usepackage{color}
\definecolor{joerg}{rgb}{1.0,0.0,0.0}

\begin{document}

\preprint{DUKE-TH-05-280}

\title{Mass and Width of the Rho Meson in a Nuclear Medium from 
       Brown-Rho Scaling and QCD Sum Rules}

\author{J\"org Ruppert}
\email{ruppert@phy.duke.edu}
\affiliation{Department of Physics, Duke University, Box 90305,  
             Durham, NC 27708, USA}
\author{Thorsten Renk}
\email{trenk@phy.duke.edu}
\affiliation{Department of Physics, Duke University, Box 90305,  
             Durham, NC 27708, USA}
\affiliation{Department of Physics, PO Box 35 FIN-40014 University of Jyv\"askyl\"a, Finland}
\affiliation{Helsinki Institut of Physics, PO Box 64 FUB-00014, University of Helsinki, Finland}
\author{Berndt M\"uller}
\email{muller@phy.duke.edu}
\affiliation{Department of Physics, Duke University, Box 90305,  
             Durham, NC 27708, USA}

\date{November 2, 2005}

\begin{abstract}
We explore the range of values of the in-medium width of a $\rho$-meson at rest which is compatibale with the QCD sum rule approach in a nuclear medium assuming vector meson dominance and a Brown-Rho scaling law of the $\rho$-meson mass with the chiral condensate. The lower and upper bounds for the in-medium width are found to be strongly increasing with the decreasing mass of the $\rho$-meson (increasing nuclear density). We also study the bounds for the in-medium width in models not satisfying the Brown-Rho scaling law. It is shown that the in-medium width depends on how rapidly the mass decreases in comparison to the change of the quark condensate. The bounds for the in-medium width increase with density only if the relative change of the quark condensate is stronger than the relative decrease in mass. 
This is important for experimental tests of the Brown-Rho scaling paradigm and other dropping $\rho$-mass scenarios.
\end{abstract}

\maketitle

\section{Introduction} 
The properties of vector mesons, such as the $\rho$-meson, are predicted to change in the presence of a dense nuclear medium due to the approach to chiral symmetry restoration and quark deconfinement at high temperature and baryon density. Reliable predictions of the nature of these in-medium modifications are thus of crucial importance for the interpretation of the experimentally measured dilepton spectra. The CERES and HELIOS-3 collaborations demonstrated that central nucleus-nucleus collisions result in a strong enhancement of the emission of low-mass dileptons as compared to scaled proton-nucleus and proton-proton collisions. These experimental findings have been interpreted as evidence for in-medium modifications of the $\rho$-meson \cite{Masera:1995ck,Agakishiev:1995xb,Agakishiev:1998vt,Lenkeit:1999xu}.

The NA60 experiment has recently presented a first high-resolution di-muon spectrum from In+In collisions at a beam energy of 158 AGeV/c \cite{Sanja}. These data make it possible to explore the in-medium change of the properties of the $\rho$-meson in unprecedented detail. In particular, the data permit the first serious test of the Brown-Rho scaling law. In this work we lay the foundations for such a test by determining the constraints on the correlation between a $\rho$-mass shift, the dropping of the absolute value of the quark condensate and a change in the width of the $\rho$-meson spectral function. 

\subsection{Brown-Rho scaling and  dropping $\rho$-mass in nuclear medium}

Brown and Rho \cite{Brown:1991kk,Brown:1995qt,Brown:2001nh,Brown:2002is} attributed the observed enhancement primarily to the gradual restoration of chiral symmetry. They proposed a scaling of the masses of the vector mesons in a medium with the quark condensate according to the relation \cite{Note1}
\begin{equation}
m^*_\rho/m_\rho \sim (\langle \bar{q}q\rangle^*/\langle \bar{q}q\rangle)^{1/2}.
\label{BR}
\end{equation}
This change will cause the maximum of the spectral function of the $\rho$-meson to be shifted to lower masses in heavy ion collisions due to the formation of a hot, dense nuclear medium and thus lead to an enhancement of lepton pair production at lower invariant masses as compared to the vacuum.

We emphasize that the scaling law (\ref{BR}) predicts a {\it specific} relation between the change of the in-medium mass and that of the absolute value of the quark condensate. Not every prediction of a dropping of the in-medium $\rho$-meson mass with nuclear density can be called Brown-Rho scaling. We find that QCD sum rule 
constraints on the correlation between a $\rho$-mass shift and the change on the width of the $\rho$-meson spectral function depend sensitively on how rapidly the $\rho$-mass changes in comparison to the change of the quark condensate in the nuclear medium. 

Our analysis is based on the framework of QCD sum rules at finite baryon density and temperature 
\cite{Shifman:1978bx}. 
The variation of the quark condensate with baryon density is expected to be stronger than with temperature (for the densities and temperatures reached in collisions at the NA60 experiment)\cite{Note0}.  Therefore we expect baryon density to be the dominant effect for Brown-Rho scaling of the masses. For simplicity we restrict our analysis to QCD sum rules finite baryon density only. 

\subsection{QCD sum rule approach} 
 
QCD sum rules have been widely employed in the analysis of the $\rho$-meson mass at finite nuclear density \cite{Hatsuda:1991ez,Asakawa:1992ht,Asakawa:1993pq,Hatsuda:1995dy,Jin:1995qg,Leupold:1997dg,Klingl:1997kf}. Universally dropping masses were supported by early studies based on this approach \cite{Hatsuda:1991ez,Hatsuda:1995dy,Jin:1995qg}. 
These found that the $\rho$-mass decreased in the presence of a nuclear medium when the $\rho$-meson spectral function was modeled as a $\delta$-function, i.~e.\  when the width of the $\rho$-meson was neglected. In \cite{Asakawa:1992ht,Asakawa:1993pq} the $\rho$-meson was studied in the QCD sum rule approach for a hadronic model which included the effects of the delta-hole polarization on the pion. On the other hand, the application of the QCD sum rule approach to an effective Lagrangian combining chiral $SU(3)$ dynamics with vector meson dominance predicted significant broadening effects but almost no mass shift of the $\rho$-meson \cite{Klingl:1997kf}.

The aim of our analysis is different. We do not want to use QCD sum rules to explore the predictions of a specific hadronic model for in-medium change of the $\rho$-meson spectral function. Rather, we follow the reasoning of Leupold {\em et al.} \cite{Leupold:1997dg} and explore the hadronic model independent constraints imposed by QCD sum rules on the mass and width of the $\rho$-meson in a dense medium. We then apply our results to dropping $\rho$-mass scenarios and derive constraints on the width of the $\rho$-meson assuming the Brown-Rho scaling law for the $\rho$-meson mass is valid. The results obtained in this way will form the basis for a consistent calculation of the dilepton spectrum in the Brown-Rho scenario and its comparison with the data in a future publication.

For simplicity, we employ the Breit-Wigner parametrization for the $\rho$-meson spectral function \cite{Leupold:1997dg}. We assume that the pole position is given by the Brown-Rho scaling law and and keep the on-shell width as a free parameter.
We also discuss in a subsection how this analysis of the compatible widths changes if the $\rho$-meson mass changes not in accordance to the Brown Rho scenario. Of course, sophisticated hadronic models can yield more complicated spectral functions, but we do not expect the sum rules to be sensitive to such details in the modeling of the spectral function. We note that our {\em ansatz} may not apply to spectral functions with peculiar features, e.~g.\ two separate pole structures. Such structures could, indeed, be generated by the coupling of the nucleon-$\rho$ channel to the $D_{13}(1520)$ resonance and are predicted to occur in a certain class of hadronic models (see e.~g.\ \cite{Peters:1997va, Post:2003hu, Leupold:2004gh}). The general approach presented here can easily be extended to constrain models of this type, as well.

We first recall the QCD sum rules in a nuclear medium for a $\rho$-meson at rest. After discussing the parametrization of the Breit-Wigner spectral function we explain how the QCD sum rules can be exploited to derive constraints on the $\rho$-meson width if a density dependence of the $\rho$-meson mass is assumed. 

Since many technical aspects of this approach have been presented in \cite{Leupold:1997dg}, we restrict ourselves here to a brief discussion.

\section{Technical Framework}

In order to derive the QCD sum rule for the vector meson channel, one improves the convergence of the operator product expansion by applying a Borel transformation. Equating the current-current correlator with its perturbative value amended by the contribution from the quark and gluon condensates, one obtains \cite{Leupold:1997dg}:
\begin{eqnarray}
\frac{1}{\pi M^2} \int\limits^\infty_0 \!\! ds \,
{\rm Im} R (s) \, e^{-s/M^2} &=&
\frac{1}{8\pi^2}\left(1+\frac{\alpha_s}{\pi} \right)
+ \frac{1}{M^4} m_q \langle \bar q q\rangle 
+ \frac{1}{24 M^4} \left\langle \frac{\alpha_s }{\pi} G^2 \right\rangle 
\nonumber \\
&& {} + \frac{A_2}{4 M^4} m_N \rho_N 
- \frac{56}{81 M^6} \pi\alpha_s \kappa \langle \bar q q\rangle^2 
- \frac{5A_4}{24 M^6} m_N^3 \rho_N. 
  \label{eq:botr}
\end{eqnarray}
This relation is taking into account condensates of six or less dimensions and is valid at twist two. We first discuss the right-hand side of (\ref{eq:botr}) where $\langle \bar q q\rangle$ denotes the quark condensate, $m_q$ the current quark mass, $\left\langle \frac{\alpha_s}{\pi} G^2 \right\rangle$ the gluon condensate and $m_N$ the nucleon mass. $\kappa$ parametrizes the deviation of the four-quark condensate from the square of the two-quark condensate. In the present study three parameter sets are used 
where it is assumed that this dependence factorizes in such a way that $\kappa$ does not depend on density.   

In our study, the medium dependence of the condensates is taken into account in leading order in the baryon density \cite{Hatsuda:1991ez}.
The terms involving the coefficients $A_2$ and $A_4$ arise from the non-scalar quark condensates (see \cite{Hatsuda:1991ez,Hatsuda:1992bv} for details). Phenomenological values for $A_2$ and $A_4$ can be derived from the quark parton distributions. 
 For the scalar condensates this dependence is: 
\begin{eqnarray}
\label{eq:cond}
  \langle \bar q q \rangle & = & 
\langle \bar q q \rangle_{\rm vac} + \frac{\sigma_N}{2 m_q} \rho_N
\nonumber \\
  \left\langle \frac{\alpha_s}{\pi} G^2 \right\rangle & = &
\left\langle \frac{\alpha_s}{\pi} G^2 \right\rangle_{\rm vac} 
- \frac{8}{9} m_N^{(0)} \rho_N \, .
\end{eqnarray}
Here $m_N^{(0)}$ is the nucleon mass in the chiral limit and $\sigma_N$ is the nucleon sigma term. We here employ three parameter sets: Set I from Hatsuda and Lee \cite{Hatsuda:1991ez}, set II from Klingl, Kaiser, and Weise \cite{Klingl:1997kf}, and set III from Leinweber \cite{Leinweber:1995fn}. The numerical values of the parameters characterizing the change of the condensates in the medium are listed in Table 1 of Leupold {\em et al.} \cite{Leupold:1997dg}. We employ these three parameter sets in order to explore the range of different predictions for the in-medium change of the condensates. As we will find, the principal findings of the present study are the same for all three parameter sets, but the quantitative predictions are somewhat dependent on the particular parametrization.

We now turn to the left-hand side of (\ref{eq:sumrule}). The dimensionless quantity $R(Q^2)=\Pi(Q^2)/Q^2$ is defined in terms of the contracted current-current correlator 
\begin{equation}
\Pi(q^2) = \int d^4x e^{iq\cdot x} \langle j_\mu(x) j^\mu(0)\rangle ,
\end{equation}
where $j_\mu$ is the electromagnetic current.
Its imaginary part can be parametrized as a sum of three contributions \cite{Hatsuda:1992bv, Jin:1995qg, Klingl:1997kf}: the contribution from the $\rho$-meson, where $S(s)$ is its spectral function assuming vector meson dominance \cite{Note2}, the perturbative continuum contribution, and the contribution from Landau damping:
\begin{eqnarray}
\label{eq:hacola}
{\rm Im} R(s) = 
\pi F \frac{S(s)}{s} \Theta(s_0-s)  
+ \frac{1}{8\pi}\left(1+\frac{\alpha_s}{\pi} \right) \Theta(s-s_0) 
+ \delta(s) \frac{\pi}{4} \rho_N \frac{1}{\sqrt{k_F^2+m_N^2}} \, ,
\end{eqnarray}
where $k_F$ denotes the nuclear Fermi momentum. Inserting this {\em ansatz} into (\ref{eq:cond}) one obtains \cite{Leupold:1997dg}:
\begin{eqnarray}
\frac{1}{M^2} \int\limits^{s_0}_0 \!\! ds \,
F \frac{S(s)}{s}\, e^{-s/M^2} & = &
\frac{1}{8\pi^2}\left(1+\frac{\alpha_s}{\pi} \right) 
\left( 1 - e^{-s_0/M^2} \right) 
- \frac{1}{4 M^2} \rho_N \frac{1}{\sqrt{k_F^2+m_N^2}} 
\nonumber \\
&& {}+ \frac{1}{M^4} m_q \langle \bar q q\rangle 
+ \frac{1}{24 M^4} \left\langle \frac{\alpha_s}{\pi} G^2 \right\rangle 
+ \frac{1}{4 M^4} m_N A_2 \rho_N 
\nonumber \\
  \label{eq:sumrule}
&& {}- \frac{56}{81 M^6} \pi\alpha_s \kappa \langle \bar q q\rangle^2 
- \frac{5}{24 M^6} m_N^3 A_4 \rho_N   \,.
\end{eqnarray}

In accordance with \cite{Leupold:1997dg} we choose the Breit-Wigner parametrization for the $\rho$-meson spectral function
\begin{equation}
\label{eq:spec}
S(s) = \frac{1}{\pi} \frac{\sqrt s \,\Gamma(s)} 
                          {(s-m^{*2}_{\rho})^2+s\,(\Gamma(s))^2} \,,
\end{equation}
where $\Gamma(s)$ has the form:
\begin{equation}
\label{eq:gammamed}
\Gamma_{\rm med}(s) = \gamma \,
\left( \frac{1- \frac{m_\pi^2 }{s}}{1- \frac{m_\pi^2 }{m^{*2}_\rho}} \right)^{1/2}
 \,\Theta(s-m_\pi^2)  \,.
\end{equation}
This parametrization of the in-medium spectral function takes into account that the lowest threshold for decay of a $\rho$-meson in the medium is given by the channel $\rho \to \pi N$. It is assumed that the dominant partial wave is  $s$-wave \cite{Leupold:1997dg}. We emphasize that this is a simplification. For example since the two-pion channel determines the decay threshold in the vacuum, one would expect that it continues to influence the shape of $\Gamma_{\rm med}(s)$ at least for low baryon densities $\rho_N$. Here we are mainly interested in the regime $\rho_N \ge 0.3 \rho_0$ where we assume that the one-pion threshold might be dominant \cite{Leupold:1997dg}. In order to explore the possible effect of a different form of the decay width on our conclusions, we have performed calculations assuming a dominant two-pion threshold behavior implying a decay width of the form \cite{Leupold:1997dg}:
\begin{equation}
\Gamma_{\pi\pi}(s) = 
\gamma \left(\frac{1-4\frac{m_\pi^2}{s}}{1-4\frac{m_\pi^2}{m^{*2}_\rho}}\right)^{3/2} 
       \Theta(s-4m_\pi^2) .
\end{equation}
These yield a compatibility corridor of widts for a given nuclear matter density that is shifted to even higher values compared with our findings for the one-pion threshold discussed below. We are thus confident that possible contributions to $\Gamma_{\rm med}(s)$ from the two-pion channel would not change our principal findings even for low nucleon densities. 

We employ standard techniques for the determination of the acceptable Borel window and the optimization of the values of $s_0$ and $F$. For details we refer the reader to the discussion in section IV. of Leupold {\em et al.} \cite{Leupold:1997dg}. We then average over the relative deviations between the two sides of (\ref{eq:sumrule}) for a given set of values of the in-medium $\rho$-meson mass $m_\rho$ and width $\gamma$ over the Borel window, schematically:
\begin{equation}
\label{eq:diffdef}
  d = \int\limits^{M^2_{\rm max}}_{M^2_{\rm min}} \!\! \frac{d(M^2)}{\Delta M^2} \, 
\left\vert 1- \frac{\rm l.h.s.}{\rm r.h.s.} \right\vert,  
\end{equation}
where $\Delta M^2=M^2_{\rm max}-M^2_{\rm min}$.

Three criteria are used to determine the compatible combinations of in-medium mass and width in a QCD sum rule approach (see e.~g.\ \cite{Leupold:1997dg}). The first criterion is that $d$ is reasonably small (we require $d \le 0.2\%$ \cite{Leupold:1997dg}). The second criterion determines the lower bound of the Borel window by demanding that the terms of the order $O(1/M^6)$ on the r.~h.~s.\ of (\ref{eq:sumrule}) do note contribute more than $10\%$ to the total value. The third criterion determines the upper bound of the Borel windows by requiring that the continuum contribution to the l.~h.~s.\ of (\ref{eq:sumrule}) is not larger than the contribution from the $\rho$ spectral function. Those conditions result in a contraction of the Borel window $\Delta M^2$ with increasing nuclear density as discussed by Jin and Leinweber \cite{Jin:1995qg}. 


\section{Results}

\subsection{Dropping $\rho$-mass in a nuclear medium in accordance with Brown-Rho scaling}
\begin{figure}[!htb]
\vspace{0.5cm}
\epsfig{file=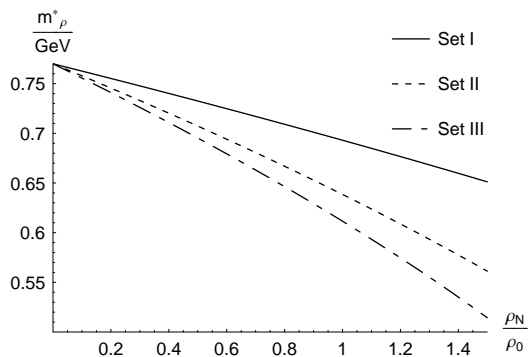, width=8 cm}
\caption{\label{figure1}The mass of the $\rho$-meson as a function of the nuclear matter density $\rho_N$ as predicted by the Brown-Rho scaling law. The solid line is for parameter set I, the dashed line for parameter set II, and the dotted dashed line for parameter set III. The parameter sets differ in their parametrization of the density dependence of condensates. For details see the main text after eq.~(\ref{eq:cond}). All three parametrizations assume a linear density dependence of the quark condensate.}
\end{figure}

After these preparations we are ready to determine the constraints from the QCD sum rule on the in-medium width in the Brown-Rho scenario. We assume the Brown-Rho scaling law for the in-medium mass $m^*_\rho$ in the form Eq. (\ref{BR}).
In Fig.~\ref{figure1} we show the implications of (\ref{BR}) for the density dependence of $m^*_\rho$, in the linear density approximation (\ref{eq:cond}) for the quark condensate and for the three parameter sets described above. 

Dynamical models of the heavy-ion collision suggest that the matter created in the In+In collisions studied by the NA60 experiment \cite{Sanja} reaches nuclear densities up to $1.3~\rho_0$. Our calculations therefore cover a density range slightly exceeding this value, namely, $0 < \rho_N < 1.5\rho_0$. The mass of the $\rho$-meson as a function of density can be approximated quite well by a linear decrease of $10.4\%$, $18.3\%$, and $22.5\%$, respectively, at normal nuclear matter density $\rho_0$ for the parameter sets I, II, and III. It is important to notice that because of (\ref{BR}) 
the corresponding decrease of the absolute value of the quark condensate  with density is even stronger, namely a linear decrease of  $18.9\%$,~$31.2\%$,~$36\%$ at normal nuclear matter density. This is significant since the relative decrease of mass and 
quark condensate determines the change in the width, see section \ref{nonBRsec}.

\begin{figure}[!htb]
\vspace{0.5cm}
\epsfig{file=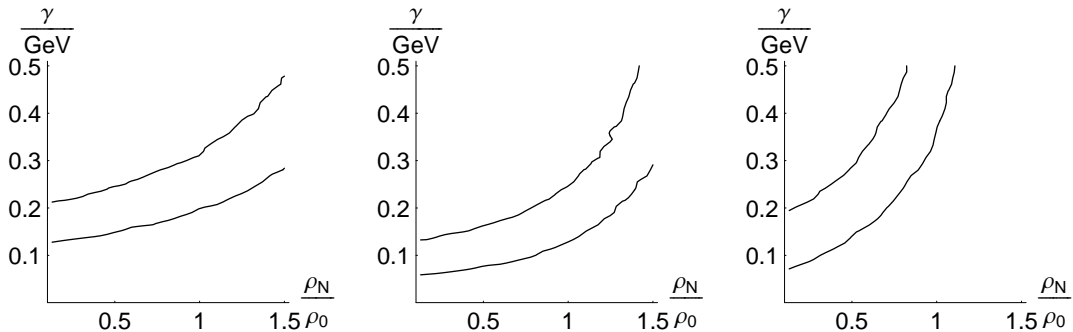, width=15 cm}
\caption{\label{figure2} The plots above show the corridors of width $\gamma$ (see Eq. (\ref{eq:gammamed})) allowed by the QCD sum rule as function of the nuclear density under the assumption of a Brown-Rho scaling law for the in-medium $\rho$-meson mass. The plots are, from left to right, for parameter set I, II, and III. The parameter sets differ in the parameterization of the density dependence of the quark condensate. For details, see the main text.}
\end{figure}

Having fixed the density dependence of the $\rho$-meson mass by means of the Brown-Rho scaling law, we can now use the QCD sum rule to calculate the allowed range of the width parameter $\gamma$ for a given nuclear density. In Fig.~\ref{figure2} we show the compatibility corridor of width for the three parameter sets. One notices that the range of allowed values moves to larger widths for increasing baryon density. The effect is much more pronounced for the parameter set III than for the other two parametrizations. In fact, set III does not yield meaningful values of the width for baryon densities above $\rho_0$ and thus does not appear suitable for a description of lepton pair production in the In+In collision system. We note that this set also yields the strongest drop of the $\rho$-meson mass with density among the three parametrizations.

Our results imply that a dropping $\rho$-mass in the framework of the Brown-Rho scaling scenario (see Eq. (\ref{BR})) must be accompanied by an increase in the $\rho$-meson width.
Figures \ref{figure1} and \ref{figure2} can be combined to associate a given baryonic density with an in-medium mass of the $\rho$-meson and a range of widths, which are compatible with the QCD sum rule approach. One observes that the modification of the width is significantly more pronounced than the quite moderate mass shift in a Brown-Rho-scaling scenario within the considered density range. This suggests that the growing in-medium width, and not the density dependent scaling of the meson mass, is the dominant effect on the $\rho$-meson spectral function and, therefore, on the dilepton spectrum in 
a scenario where Brown-Rho scaling holds. In particular, the enhancement of dilepton emission for invariant masses below the minimal $\rho$-meson mass reachable in the Brown-Rho scaling scenario (about 555 MeV at $\rho_N = 1.3~\rho_0$ for parameter set III assuming the dominance of density over temperature caused mass shifts), must be attributed to the increased broadening of the spectral function in the medium.

\subsection{Non-Brown-Rho scaled dropping $\rho$-mass in a nuclear medium}
\label{nonBRsec}

In this subsection we study the question what happens if the $\rho$-meson mass  in the medium does not scale 
in accordance with the Brown-Rho scaling law. Since models predicting a raising $\rho$-mass (see e.g. \cite{Pisarski:1994yp}) face a hard challenge by the experimentally observed strong enhancement of the emission of  dileptons below the vacuum $\rho$ mass, we restrict our discussion here to dropping $\rho$-meson masses in the nuclear medium.

\begin{figure}[!htb]
\vspace{0.5cm}
\epsfig{file=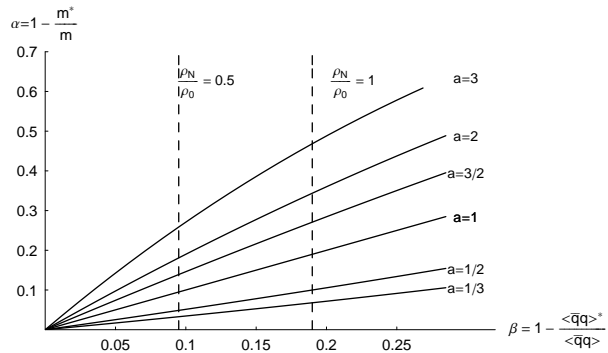, width=8cm}
\caption{\label{figureNON1} 
The relative change of the $\rho$-mass is in a function of density and via Eq. (\ref{eq:cond}) of  the relative change of the quark condensate. The vertical dashed lines indicate $0.5$ and $1$ times normal nuclear matter density, respectively. The QCD sum rule method provides upper and lower bounds for the width $\gamma$ for a given density if the in-medium mass change is given.  We consider a generalized scaling law, cmp. Eq. (\ref{general}), in order to give an impression how the width changes depending on how the in-medium mass changes with density, see Fig. \ref{figureNON2}.\,The case $a=1/2$ corresponds to the Brown-Rho scaling law, cmp. Eq. (\ref{BR}) and \cite{Note1}. }
\end{figure}

\begin{figure}[!htb]
\vspace{0.5cm}
\epsfig{file=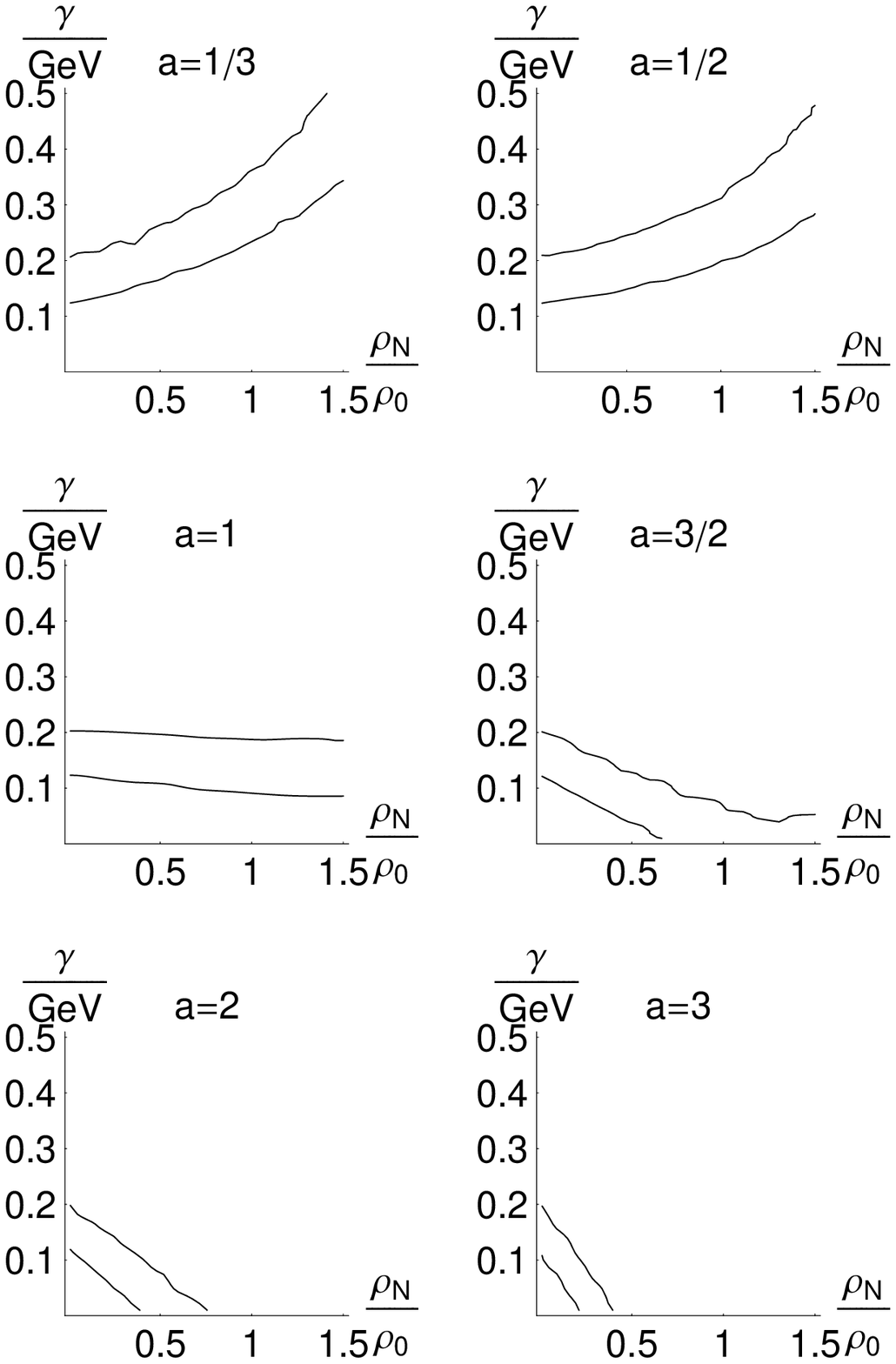, width=10cm}
\caption{\label{figureNON2} 
The $\rho$-mass is in general a function of density.  The QCD sum rule method provides upper and
lower bounds for the width $\gamma$ (see Eq. (\ref{eq:gammamed})) for a given density if the in-medium mass change is given. 
In order to give an impression how the width changes depending on how the in-medium
mass changes with density a generalized scaling law is considered, cmp. Eq. (\ref{general}). The case $a=1/2$ corresponds to the Brown-Rho scaling law, cmp. Eq. (\ref{BR}) and \cite{Note1}. Going to higher values of $a$ one sees that a assuming a faster dropping of mass with density eventually turns the increase of the bounds of the width into a strong decrease with density. }
\end{figure}

In Fig. \ref{figureNON1} we show the plane of the relative changes of the $\rho$-mass and quark condensate. 
Since we are here mainly interested in a qualitative discussion, we focus only on one parameter set, namely set I, for the QCD sum rule analysis. The leading-order change of the quark condensate with baryon density is given by Eq. \ref{eq:cond}. Therefore the vertical dashed  lines indicate the constant
nuclear density. Every model prediction 
of the $\rho$-meson in-medium mass  corresponds in this framework to a prediction of how $m^*$ changes
as a continuous function of $\langle \bar{q}q \rangle^*$, starting at the vacuum values of the quantities at zero density. We define two quantities $\alpha=1-m^{*}/m$ and $\beta=1-\langle \bar{q} q\rangle^{*}/\langle \bar{q} q\rangle$ that express the relative change of the mass and condensate with density, respectively.
  
Our QCD sum rule analysis allows for the calculation of a two-dimensional surface above this plane indicating the
lower ($\gamma_{\rm min}(\alpha, \beta)$) and upper ($\gamma_{\rm max}(\alpha, \beta)$) bounds for the compatible width $\gamma$. Since the full surfaces are cumbersome to discuss,
we restrict ourselves here to cuts through these surfaces corresponding to different
scaling laws of the $\rho$-mass with density. 

In order to study how the prediction of the lower and upper bounds of the width vary depending
on how the $\rho$-mass changes with density, we consider
scaling laws of the general form:
\begin{eqnarray}
\label{general}
\frac{m_\rho^*}{m_\rho}=\left(\frac{\langle \bar{q} q \rangle^*}{\langle \bar{q} q\rangle}\right)^{a}.
\end{eqnarray}
The case $a=1/2$ corresponds to the Brown-Rho scaling law \cite{Note1}. 
The solid lines in Fig. \ref{figureNON1} indicate scaling laws with $a=1/3,\,1/2,\,1,\, 3/2,\,2,\,3$, respectively: 
the higher the value of $a$ the stronger is the drop of the $\rho$-mass with density.

Fig. \ref{figureNON2} shows the constraints on the lower and upper bound of the width for the different
scaling laws. One realizes that for a stronger drop of $\rho$-mass with density (higher $a$) the 
band of compatible width is shifted to lower masses.
Especially if the relative drop of the $\rho$-mass is 
faster than the relative decrease of the absolute value of the condensate ($a>1$) then the band of compatible 
width is dropping considerably with density. Whereas if the situation is reversed ($a<1$) the band of compatible width
considerably rises with density -- as already predicted for the case of the Brown-Rho scaling 
scenario in the section \label{BRsec}.

This analysis also explains an ostensible discrepancy from comparisons of our
QCD sum rule analyses in the last subsection with earlier work on QCD sum rules, 
which is most significant in the case where one assumes that 
the meson spectral function can be described by a $\delta$-function, i. e. vanishing width.
QCD sum rule analysis showed that the mass of the $\rho$-meson drops fast with
density using this ansatz, see e.g. \cite{Hatsuda:1991ez,Hatsuda:1995dy,Jin:1995qg,Asakawa:1993pq}. The density dependence of the  in-medium meson mass  is also almost the same if one calculates the in-medium mass from the change of the bare $\rho$-meson mass using an an effective meson Lagrangian, see Fig. 1 in \cite{Asakawa:1993pq}. One finds a considerable dropping width of the spectral function for densities $>\rho_0$ and a dropping mass. 
Since  in \cite{Asakawa:1993pq} the authors employ the (slightly modified) parameter set
of Hatsuada and Lee \cite{Hatsuda:1991ez} in their analysis the in-medium mass change is not according to Eq.
 (\ref{BR}), instead the mass drops much faster than one would expect from the Brown-Rho scaling law. 
 
This situation is in qualitative agreement with our analysis. In scenarios where the mass drops much faster 
than one would expect from the Brown-Rho scaling law, e.g. in the cases $a>1$, one finds a strongly dropping
band of compatible width with density.  In that sense there is no discrepency between our 
analysis and the statement that a dropping in-medium vector 
meson mass with density does not necessarily lead to an increase 
in the compatible width in a QCD sum rule analysis. 

Although scenarios with a strongly dropping $\rho$-mass and therefore strongly dropping width cannot in general be ruled out solely on the basis of QCD-sum rule analysis, we emphasize 
such scenarios are also not supported by hadronic model calculations. With the two-pion decay being the dominant decay channel of the $\rho$-meson in the vacuum, modifications of the pion in a nuclear medium must influence the $\rho$-meson spectral function \cite{Rapp:1995zy,Rapp:1999ej,Ruppert:2004yg}. Furthermore, interactions of the $\rho$-meson with surrounding nucleons are also known to have a substantial effect on the $\rho$-meson spectral function \cite{Rapp:1999ej}. Both mechanisms have been studied in hadronic models and can be expected to result in a significant broadening of the $\rho$-meson's spectral function in a dense, baryon-rich hadronic medium. We also know that the NA60 experiment's dilepton data are not in agreement with a
strongly dropping mass assuming essentially no broadening \cite{Sanja}.

\section{Summary and conclusions}

Let us summarize our findings in the combined Brown-Rho QCD sum rule approach as presented here. Brown-Rho scaling implies a decrease of the mass of the $\rho$-meson of about $10 - 22.5\%$ at normal nuclear matter density depending on the parametrization of the density dependence of the quark condensate. The vector meson QCD sum rule demands that the medium induced change in the mass is accompanied by a considerable increase in the width of the $\rho$-meson in a Brown-Rho scaling scenario. We have calculated the range of widths that is compatible with the QCD sum rule for a given nuclear density under the assumption of the Brown Rho scaling law. The effect is similar for all three parametrizations of the density dependence of the QCD condensates considered here, but the quantitative results differ between the parametrizations. All parametrizations predict a substantial broadening as the $\rho$-meson mass drops with increasing nuclear density according to the Brown-Rho scaling law. We pointed out that this result is not general for all models that predict a dropping in-medium $\rho$-mass not in accordance with (\ref{BR}). Especially if the relative drop of the in-medium mass with density is stronger
than the relative drop of the absolute value of the quark condensate then QCD sum rule analysis predicts
a strong decrease in the compatible widths with density, but those scenario face hard challenges from 
experimental data and hadronic models indicating a raising width with density.

Our results have significance for the theoretical analysis of the new 
high-precision di-muon data from the NA60 experiment. They show that an analysis which only considers 
the effect of the Brown-Rho scaling law on the $\rho$-meson mass, but does not take into account the 
collision broadening in the medium, is inconsistent with QCD. The question to be answered by the experiment 
and its theoretical analysis is not whether the $\rho$-meson mass drops according to Brown-Rho scaling {\em or} its width increases in the dense medium, but whether the data require a dropping meson mass {\em in addition} to the broadening of the spectral function or not. We intend to report calculations of the dilepton spectrum in a dynamical model of the nuclear collision and a comparison with the NA60 data elsewhere in the near future.

\section*{Acknowledgements}
We are indebted to Stefan Leupold for helpful discussions and for providing us with a copy of his numerical code. We thank D.~D.~Dietrich for discussions. This work was supported, in part, by U.~S.~Department of Energy under grant DE-FG02-05ER41367. JR and TR thank the Alexander von Humboldt Foundation for support as Feodor Lynen Fellows. 

\bibliography{u4}

\end{document}